\RequirePackage{fix-cm}

\documentclass{svjour3}     

\smartqed  
\usepackage{graphicx}
\usepackage{amsmath,amssymb}
\usepackage{amsbsy}
\usepackage{bm}
\usepackage{float}

\usepackage[linktocpage]{hyperref}
\usepackage[usenames,dvipsnames]{color}
\usepackage{pifont}
\definecolor{darkred}{rgb}{0.5,0,0}
\definecolor{darkgreen}{rgb}{0,0.5,0}
\definecolor{darkblue}{rgb}{0,0,0.5}
\definecolor{prussian}{rgb}{0.0, 0.19, 0.33}
\definecolor{richelectricblue}{rgb}{0.03, 0.57, 0.82}
\definecolor{teal}{rgb}{0.0, 0.5, 0.5}
\definecolor{mediumseagreen}{rgb}{0.24, 0.7, 0.44}
\definecolor{lust}{rgb}{0.9, 0.13, 0.13}
\definecolor{ballblue}{rgb}{0.13, 0.67, 0.8}
\definecolor{darkcyan}{rgb}{0.0, 0.55, 0.55}
\definecolor{mountainmeadow}{rgb}{0.19, 0.73, 0.56}
\definecolor{palecarmine}{rgb}{0.69, 0.25, 0.21}
\definecolor{richcarmine}{rgb}{0.84, 0.0, 0.25}
\definecolor{tangelo}{rgb}{0.98, 0.3, 0.0}
\definecolor{venetian}{rgb}{0.784,0.031,0.082}
\definecolor{bdfrance}{rgb}{0.192,0.549,0.906}
\hypersetup{colorlinks=true, citecolor=venetian, linkcolor=bdfrance, urlcolor=lust}

\newcommand{\be}{\begin{equation}}
\newcommand{\ee}{\end{equation}}
\newcommand{\bear}{\begin{eqnarray}}
\newcommand{\eear}{\end{eqnarray}}

\newcommand{\p}{\prime}

\newcommand{\nn}{\nonumber}

\newcommand{\cL}{{\cal L}}
\newcommand{\cE}{{\cal E}}

\newcommand{\cT}{{\cal T}}

\newcommand{\cI}{{\cal I}}
\newcommand{\cJ}{{\cal J}}
\newcommand{\cH}{{\cal H}}

\newcommand{\rc}{{\rm c}}

\newcommand{\bx}{\mathbf{x}}
\newcommand{\bv}{\mathbf{v}}
\newcommand{\bxi}{\boldsymbol{\xi}}
\newcommand{\bom}{\boldsymbol{\omega}}
\newcommand{\bOm}{\boldsymbol{\Omega}}
\newcommand{\bcJ}{\boldsymbol{{\cal J}}}
\newcommand{\bP}{\boldsymbol{{\cal P}}}
\newcommand{\bnabla}{\boldsymbol{\nabla}}

\journalname{Foundations of Physics}

\begin{document}

\title{A Machian reformulation of Quantum Mechanics}

\author{Kostas Glampedakis}

\institute{Kostas Glampedakis \\ 
Departamento de F\'isica, Universidad de Murcia, Murcia, E-30100, Spain \\ 
\email{kostas@um.es}}

\date{Received: date / Accepted: date}

\maketitle

\begin{abstract}
The widely known but also somewhat esoteric Mach principle envisages a fully relational formulation of 
physical theories without any reference to a concept of `absolute space'. When applied to classical mechanics, 
under the guise of an extended symmetry group, this procedure  is known to lead to an equation of motion with 
inertial-like forces that are sourced by the mass distribution of the system itself.  In this paper we follow a similar 
procedure and reformulate the Schr\"odinger equation of non-relativistic quantum mechanics in a fully Machian 
way. Just like its classical counterpart, the resulting quantum theory is fully relational in the positions and momenta of the 
bodies comprising a given physical system, leaving no room for the notion of absolute space.

\end{abstract}


\section{Introduction}
\label{sec:intro}

The late 19th century principle formulated by Ernst Mach~\cite{MachOriginal} was born out of the criticism to Newton's famous 
buck experiment and the associated concept of absolute space in classical mechanics (for modern reviews of Mach's principle 
see Refs.~\cite{MachBook,review}). The same principle is known to have played an important role in the early-stage development 
of Einstein's General Relativity (GR) although, as it turned out, the final theory failed to be fully Machian. 

Among the subsequent discussions of the principle and of the closely related concept of the `origin of inertia' stand out the 
phenomenological  model of Sciama~\cite{Sciama53}, Lynden-Bell's attempt to build a Machian extension of 
GR~\cite{LyndenB67} and Barbour \& Bertotti's first implementation of Mach's principle as a dynamical theory~\cite{BB82}.
In parallel, a significant body of work has studied the possibility of a Machian origin of the 
relativistic dragging of inertial frames effect; early calculations based on a rotating shell model appear to lend support to 
this connection~\cite{Brill66} but later work has claimed frame dragging to be anti-Machian~\cite{Rindler94} (a modern review 
on this subject can be found in Ref.~\cite{Pfister07}). 
Although never truly achieving front-stage actor status, Mach's principle and the concepts associated with it are still subjects
of investigation in cosmology and gravitational physics (for a very recent example see~\cite{astro21}) and the question
to what extent our Universe is `Machian' is still an open one.

Mach's principle is far easier conceptualised and implemented within non-relativistic physics. The reformulation of classical 
mechanics without an absolute space reference frame by Lynden-Bell \& Katz~\cite{LyndenB95} provides a prime example 
of this assertion. This theory (which is equivalent to that of Ref.~\cite{BB82}) is designed to be invariant under a larger group of symmetries than Newtonian mechanics and 
as a consequence it is fully \emph{relational}, that is, its Lagrangian depends only on the relative positions and velocities of the 
bodies that make up the `universe' in this model. More recent work on the same model~\cite{Ferraro16} has established 
that the familiar inertial forces of Newtonian mechanics are determined by the global properties of the system itself 
rather than the motion with respect to an inertial frame of `fixed stars'; this is the very essence of Mach's principle.

Although Mach's principle has  traditionally  been considered in relation with inertial forces, relativistic  frame dragging
and cosmology,  there is no deep reason for it to be confined within the realm of classical physics. After all, standard
non-relativistic quantum mechanics (QM) is formulated in the same inertial frame  `absolute space' as Newtonian mechanics
and shares the same group of space-time symmetries. To the best of our knowledge no such Machian formulation of QM
exists in the literature and it is the purpose of this paper to develop one. Our approach is based on the extension 
of the classical model of  Lynden-Bell, Katz \& Ferraro~\cite{LyndenB95,Ferraro16}  and the final product is the derivation of 
a fully relational Schr\"odinger equation.  This is the only -- but crucial -- Machian modification of standard QM.

The rest of the paper is organised as follows. Section~\ref{sec:classical} sets the stage by summarising the relational-Machian 
model of classical mechanics developed in Refs.~\cite{LyndenB95,Ferraro16}. Section~\ref{sec:nospace} provides the link between
the extended symmetry group of the new framework and its Machian character. In Section~\ref{sec:newL} the relevant Lagrangian is 
constructed and a new `Mach-Newton' equation of motion is derived. Some implications of this  equation are discussed in 
Section~\ref{sec:implications}. The QM portion of the paper can be found in Section~\ref{sec:Hamiltonian} 
(construction of the classical relational Hamiltonian), Section~\ref{sec:operators} (definition of quantum operators) and 
Section~\ref{sec:MS} (derivation of a `Mach-Schr\"odinger' equation). Our concluding remarks can be found in 
Section~\ref{sec:conclusions}.
Throughout the paper the indices $\{i,j\}$ (with or without primes) are exclusively used as `body labels'. All other latin indices 
represent vector/tensor components. Calligraphic symbols refer to parameters in the system's center of mass frame. 
An overdot (hat) denotes a time derivative (quantum operator). 


\section{Setting the stage: classical mechanics without absolute space}
\label{sec:classical}

\subsection{From absolute space to relational space}
\label{sec:nospace}

This section summarises previous work by Lynden-Bell \& Katz~\cite{LyndenB95} (as well as its later extension by Ferraro~\cite{Ferraro16}) 
who were among the first to develop a rigorous and fully Machian reformulation of classical Newtonian mechanics without any tethers to the concept 
of absolute space. As pointed out in the introduction, this model serves as the necessary stepping stone towards a Machian treatment of QM
and for that reason we believe it should be discussed here in some detail. 

The `universe' in this model comprises  a collection of $N$ point-like bodies of mass $m_i$ and position $\bx_i (t)$ (where $i=1,2, ...N$). 
Each pair $ij$ of bodies is assumed to interact via a potential,
\be
V_{ij} = V_{ij} (\bx_{ij} ), \qquad
\ee
which is only a function of the pair's \emph{relative} position $ \bx_{ij} \equiv \bx_i - \bx_j$. For the specific example of gravitationally interacting
bodies we would have
\be
V =  \sum_{i<j} \sum_{j=1}^{N} V_{ij}   \equiv - \sum_{i<j} V_{ij},
\ee
with
\be
V_{ij} (\bx_{ij} )= -  \frac{ G m_i m_j}{x_{ij}}.
\ee
The system's Lagrangian is defined in the usual way,
\be
L = T -V =  \frac{1}{2} \sum_i m_i \bv_i \cdot \bv_i - V(\bx_{ij}). 
\label{L1}
\ee
where $\bv_i = \dot{\bx}_i$. 
This functional encapsulates the familiar symmetries attached to Newtonian mechanics:
(i) invariance with respect to time-independent rigid spatial translations and rotations,
\be
\bx_i \to \bx_i + \bxi, \qquad \bx_i \to \bx_i + ( \mathbf{A} \times \bx_i ),
\ee
where $\mathbf{A}$ is a constant vector along the axis of rotation and (ii) invariance with respect to the Galilean transformation (GT),
\be
t \to t + \epsilon, \qquad \bx_i \to \bx_i + \mathbf{V} t,
\ee
where $\epsilon$ and $\mathbf{V}$ are constants. Indeed, the privileged family of inertial frames defined by the GT represents 
the modern incarnation of Newton's `absolute space' where the theory is supposed to be valid.

Newtonian mechanics can be converted to a fully relational theory (i.e. free of the concept of absolute space) provided it is defined via 
a Lagrangian that is invariant with respect to a wider set of gauge symmetries, namely, infinitesimal time-dependent rigid translations 
(also known as `extended GT' in the literature~\cite{Green01}) and rigid rotations, 
\begin{align}
& \bx_i \to \bx_i + \bxi (t) ~ \Rightarrow ~ \bv_i \to \bv_i + \dot{\bxi},
\label{gauge1}
\\
& \bx_i \to \bx_i + \left ( \mathbf{a}(t) \times \bx_i \right ) ~\Rightarrow ~ \bv_i \to \bv_i + (\bom \times \bx_i) + ( \mathbf{a} \times \bv_i ),
\label{gauge2}
\end{align} 
where $  \bom (t)  \equiv \dot{\mathbf{a}} $. These transformations only affect the kinetic energy $T$ in~\eqref{L1}; 
the transformation~\eqref{gauge1} produces a variation 
\be
\delta T =  \dot{\bxi} \cdot \mathbf{P},
\ee
where $\mathbf{P}$ is the system's total momentum. This variation can be eliminated if we instead work with the 
kinetic energy in the center of mass (CM) frame,
\be
\cT \equiv T - \frac{1}{2} M u_c^2,
\label{Tdecomp}
\ee
where $M= \sum_i m_i$ is the total mass and
\be
\mathbf{u}_\rc = \frac{1}{M} \sum_i m_i \bv_i = \frac{1}{M} \mathbf{P},
\ee
is the CM velocity. The new kinetic energy is invariant, $\delta \cT=0$, with respect to~\eqref{gauge1} and can
be rewritten in the equivalent forms~\cite{LyndenB95,Ferraro16}
\be
\cT =  \frac{1}{2} \sum_i m_i  |\bv_i - \mathbf{u}_\rc |^2 = \frac{1}{2} \sum_{i <j} m_{ij} v_{ij}^2.
\label{Tc1}
\ee
Notice that $\cT$ resembles the potential energy term, being a double sum and depending on the relative velocities 
$\bv_{ij} \equiv \bv_i - \bv_j$ instead of the original absolute velocities. At the same time the single-body masses have been 
replaced by the pair masses  $m_{ij} \equiv m_i m_j /M $ which are a generalisation of the familiar two-body system's reduced mass. 

The kinetic energy $\cT$ is still not invariant with respect to the second transformation~\eqref{gauge2}; the associated 
variation is found to
be~\cite{LyndenB95,Ferraro16},
\be
\delta \cT = \bom \cdot \bcJ,
\ee
where $\bcJ$ is the system's total angular momentum with respect to the CM frame. This can be written in the three equivalent forms,
\be
 \bcJ \equiv  \sum_i m_i (\bx_i \times \bv_i ) -  (\bx_\rc \times \mathbf{P} )  =  \sum_i m_i ( \bx_{ic} \times  \bv_{ic} ) 
 =   \sum_{i <j} m_{ij} ( \bx_{ij}  \times \bv_{ij} ), 
\label{J1}
\ee
where we have defined the relative to the CM position and velocity,
\be
\bx_{ic} \equiv \bx_i - \bx_\rc, \qquad \bv_{ic} \equiv \bv_i -\mathbf{u}_\rc.
\ee
In particular, according to the last equation in~\eqref{J1} we have $\bcJ= \bcJ (\dot{\bx}_{ij}, \bx_{ij})$.

As in the previous case with the  modification \eqref{Tdecomp}, $\delta \cT$ can be eliminated after removing a CM-related 
rotational term from the kinetic energy.  The new, fully gauge-invariant, kinetic energy is found to be
\be
\cT^* = \cT - \frac{1}{2} \bcJ \cdot \bOm =  \frac{1}{2} \sum_{i <j} m_{ij} v_{ij}^2 - \frac{1}{2} \cI_{nk}^{-1} \cJ_n \cJ_k.
\label{Tstar}
\ee
The symmetric tensor $\cI_{kn} $ is the system's intrinsic moment of inertia with respect to the CM. This is,
\be
\cI_{nk} = \sum_i m_i  \left ( x_{ic}^2 \delta_{nk} - x^n_{ic} x^k_{ic} \right)
=  \sum_{i <j} m_{ij} \left ( \, x_{ij}^2 \delta_{kn} -x_{ij}^k  x_{ij}^n \,\right ).
\label{Ic1}
\ee
As evident from the last equation this moment of inertia is a fully relational quantity,  $\cI_{nk} = \cI_{nk} (\bx_{ij})$.
As a result of its definition $\cI_{nk}$  is related in the usual way to the absolute space moment of inertia $I_{nk}$,
 \be
I_{nk} =  \sum_i m_i \left ( x_i^2 \delta_{nk} - x^n_i x^k_i \right ) = \cI_{nk} + M \left ( x_\rc^2 \delta_{nk} - x_\rc^n x_\rc^k \right ).
\label{Ic2}
\ee
The vector $\bOm = \bOm (\bx_i, \bv_i)$ in~\eqref{Tstar} is given by, 
\be
 \cJ_n = \cI_{nk} \Omega_k ~\Rightarrow ~ \Omega_k = \cI_{nk}^{-1} \cJ_n,
\ee
from which it follows that $\bOm$ is the system's angular velocity with respect to the CM.

\subsection{The new relational Lagrangian \& the `Mach-Newton' equation of motion}
\label{sec:newL}

We can now assemble the previous results and define a fully relational Lagrangian that describes the present Machian 
reformulation of Newtonian mechanics. This Lagrangian, first derived in Ref.~\cite{LyndenB95}, is:
\be
\cL (\bx_{ij},\dot{\bx}_{ij}) \equiv \cT^* - V =  \frac{1}{2} \sum_{i <j} m_{ij} \dot{\bx}_{ij}^2 - \frac{1}{2} \cI_{nk}^{-1} \cJ_n \cJ_k - V(\bx_{ij}).
\label{L3}
\ee
The same Lagrangian can be written in several equivalent ways. For example, two forms that show explicitly the CM's central role
in this Machian framework are,
\be
\cL = L  - \left ( \frac{1}{2} M u_\rc^2 + \frac{1}{2} \bcJ \cdot \bOm \right )
= \frac{1}{2} \sum_i m_i \left  | \bv_{ic}  - \bOm \times \bx_{ic}  \right  |^2 -  V(\bx_{ij}).
\label{L2}
\ee
The $i$-body's equation of motion can be derived from the above Lagrangian in the standard way; working with the conjugate 
parameters $\{ \bx_{ic}, \dot{\bx}_{ic} \}$ we first find the canonical momentum
\be
\bP_{i} = \frac{\partial \cL}{\partial \dot{\bx}_{ic} } =  m_i  \left ( \bv_{ic} - \bOm \times \bx_{ic}  \right ).
\label{Pcanon2}
\ee
The corresponding Euler-Lagrange equation is,
\be
\frac{d \bP_{i} }{dt} = \frac{\partial \cL}{\partial \bx_{ic}},
\ee
and leads to the equation of motion
\be
 m_i \ddot{\bx}_{ic} =  -\bnabla_{ic} V +  m_i \left ( \dot{\bOm }\times  \bx_{ic}  +  \bOm\times  \dot{\bx}_{ic} \right )
-  \frac{1}{2}\bnabla_{ic} \left ( \bcJ \cdot \bOm \right ),
\label{eom1}
\ee
where the gradient is taken with respect to $\bx_{ic}$. After manipulating the last term this becomes,
\be
m_i  \ddot{\bx}_{ic}  = -\bnabla_{ic} V  +  m_i \left [  \dot{\bOm} \times  \bx_{ic}  + 2  \bOm \times \dot{\bx}_{ic} 
 - \bOm \times \left ( \bOm \times \bx_{ic} \right ) \,\right ].
 \label{eom2}
\ee
We would have obtained an identical result had we used the variables $\{ \bx_i, \dot{\bx}_i \}$ (e.g. note that 
 $\bnabla_{ic} V (\bx_{ij}) = \bnabla_{ic} V (\bx_{ic}-\bx_{jc}) $ is equal to $ \bnabla_{i} V (\bx_{ij}) = \bnabla_{i} V (\bx_i-\bx_j)$). 

The `Mach-Newton' equation~\eqref{eom2} is the central result of this section and was previously obtained in~\cite{Ferraro16}. 
It is structurally identical to the textbook Newton's law in a non-inertial frame attached to the system's CM and rotating with angular 
frequency $-\bOm$ with the usual inertial forces present. However, appearances are deceiving because these forces are, in fact, 
generated by the spatial distribution and motion of the rest of the  `universe'  (through the function $\bOm$).  This property of the 
new theory is highly non-trivial because these Machian terms are present even when there is \emph{no} direct interaction, $V=0$, 
between the bodies; one could claim that these terms represent  the `backreaction' of space to the presence of matter. 

All distances and velocities in Eq.~\eqref{eom2} are measured relative to the CM. The motion of the CM itself is left unconstrained 
since the new theory does not supply an equation of motion for it. This is in contrast to standard Newtonian mechanics where the CM's 
motion is prescribed relative to `absolute space'.  The true Newtonian limit of ~\eqref{eom2} can be easily identified from~\eqref{L2} 
and corresponds to $\dot{\mathbf{u}}_\rc = \bOm =0$, that is, uniformly moving CM and vanishing total angular momentum in the CM frame. 
The privileged frames where these conditions are met can be rightly called `Newtonian frames' and  their definition implies 
that these frames are related to each other through time-independent rigid translations and rotations (plus the usual GT).


\subsection{A gravitational bucket experiment}
\label{sec:implications}

No discussion of a Machian theory  would be complete without a remark on the famous rotating bucket experiment. 
In Ref.~\cite{Ferraro16} the `bucket' is modelled as a binary system of two point masses $m$ moving in a circular orbit of radius 
$R$ about their CM. As a proxy for the `fixed stars' this model assumes a distant spherical shell centered at the binary's CM. 

Following Ref.~\cite{Ferraro16}, we work in the binary's corotating frame and find that Eq.~\eqref{eom2} reduces to
the following balance between the gravitational and `centrifugal' forces (the net gravitational force 
exerted by the shell on the binary is  of course equal to zero), 
\be
 \frac{G m^2}{4R^2}   = m \Omega^2 R.
 \label{bucket0}
\ee
The shell's angular momentum is $\bcJ = I_0 \bOm_{\rm b} $ where $I_0$ is the shell's isotropic moment of inertia
and $ \bOm_{\rm b} =\Omega_b  \mathbf{\hat{z}}$ is the binary's orbital frequency. At the same time $ \cJ_n = \cI_{nk} \Omega_k $
and we find,
\be
\bOm = \frac{I_0}{\cI_{zz}}  \bOm_{\rm b}.
\ee
The principal axis component $\cI_{zz}$ can be obtained from \eqref{Ic2}, leading to $\cI_{zz} = I_0 + 2 m R^2$. Inserting this 
in~\eqref{bucket0}
\be
\frac{G m}{4R^2} =  \Omega_{\rm b}^2 R \left (1+ \frac{2m R^2}{I_0} \right )^{-2} 
\label{bucket1}
\ee
This key expression illustrates how the standard centrifugal acceleration is related  to the large-scale distribution of matter. 
Once the shell is removed, $I_0 \to 0$, the inertial force vanishes altogether in agreement with Mach's principle. 
The more realistic situation is the limit $2mR^2/I_0 \ll 1$ which leads to the familiar centrifugal force of Newtonian mechanics. 

An alternative interpretation of \eqref{bucket1} presents itself if we write it as the standard Kepler's third law with an
effective non-local gravitational constant 
\be
G_{\rm eff} \equiv G  \left (1+ \frac{2m R^2}{I_0} \right )^{2}. 
\label{bucket2}
\ee
As any practitioner of GR would instantly spot, this relation signals a violation of the weak equivalence principle. 
The loss of the universality of free fall in a gravitational field is a completely general property of the new Machian 
theory and a clear demonstration of its non-GR character. This can be easily seen by writing down the equation of 
motion~\eqref{eom2} for the case of gravitationally interacting bodies,
\be
\ddot{\bx}_{ic}  = -\sum_{j\neq i} \frac{G m_j}{x_{ij}^3} \bx_{ij}  +   \dot{\bOm} \times  \bx_{ic}  + 2  \bOm \times \dot{\bx}_{ic} 
-  \bOm \times \left ( \bOm \times \bx_{ic} \right ).
\ee
Unlike its Newtonian counterpart, this equation does depend on the body's mass $m_i$ as it enters the global variable $\bOm$.


\section{Quantum mechanics `without absolute space'}

\subsection{The classical Hamiltonian}
\label{sec:Hamiltonian}

The next order of business following the successful Machian reconstruction of classical mechanics is the 
formulation of a non-relativistic QM theory without the concept of absolute space.

The first step towards that goal is the construction of an appropriate classical Hamiltonian. Starting from the definition,
\be
\cH \equiv \sum_i \bP_{i} \cdot \bv_{ic} -\cL(\bv_{ic},\bx_{ic}),
\label{Ham1}
\ee
we can subsequently manipulate the first term,
\be
 \sum_i  \bP_i \cdot \bv_i = \sum_i \frac{{\cal P}_i^2}{m_i} + \mathbf{u}_\rc \cdot \sum_i \bP_i
 + \bOm \cdot \sum_i  ( \bx_{ic} \times \bP_i ).
\ee
It is straightforward to verify that the canonical momenta obey the conditions,
\be
\sum_i \bP_i = 0, \qquad  \sum_i  \left (  \bx_{ic} \times \bP_i \right ) = 0.
\label{Pconds}
\ee
These carry a clear physical meaning: the system's total canonical momentum and angular momentum 
(with respect to the CM) are equal to zero. As a result of these conditions the Hamiltonian takes the
 simple `single particle' form
\be
\cH = \sum_i \frac{{\cal P}^2_{i}}{2m_i} + V(\bx_{ij}) 
=   \sum_i \frac{ | \mathbf{p}_{ic} - m_i \bOm \times \bx_{ic}|^2}{2m_i} + V(\bx_{ij}),
\label{Ham2}
\ee
where $\mathbf{p}_{ic} = m_i \bv_{ic}$ is the $i$-body's kinematical momentum relative to the CM. 
As expected, $\cE = \cH$ represents the system's conserved energy:
\begin{align}
\frac{d\cH}{dt} &=  \sum_i \left ( \dot{\bP}_i \cdot \bv_{ic} + \bP_i \cdot \dot{\bv}_{ic} -  \frac{\partial \cL}{\partial \bx_{ic}} \cdot \bv_{ic} 
- \frac{\partial \cL}{\partial \bv_{ic}} \cdot \dot{\bv}_{ic} \right )
\nn \\
&= \sum_i \left ( \dot{\bP}_i -  \frac{\partial \cL}{\partial \bx_{ic}} \right )  \cdot \bv_{ic} =0.
\end{align}
It is worth pointing out that, modulo the previous subtraction of the rotational term $ \bcJ \cdot \bOm/2$ in~\eqref{Tstar}, 
the Hamiltonian~\eqref{Ham2} looks structurally identical to the classical mechanics Hamiltonian in a rotating frame 
(e.g. see Ref.~\cite{Gulshani78}). In reality the two cases are physically distinct due to the presence of the Machian
function $\bOm (\bx_{ij}, \bv_{ij})$ in~\eqref{Ham2}.


\subsection{The position and momentum operators}
\label{sec:operators}

Having at our disposal the classical relational Hamiltonian~\eqref{Ham2} we are in a position to build a relational theory of QM.
For simplicity, we will work in the position representation; following the established textbook approach~\cite{Messiah} 
we first introduce the usual operators for the fundamental triad of position, energy and (kinematical) momentum: 
\be
\hat{\bx}_{ic} = \bx_{ic}, \qquad \hat{\cE} = i\hbar \frac{d}{dt}, \qquad \mathbf{\hat{p}}_{ic} = - i\hbar \bnabla_{ic}.
\ee
These operators obey the fundamental commutation relation, 
\be
\left [x_{ic}^k, \hat{p}_{jc}^n \right ] = i \hbar\, \delta_{nk} \delta_{ij}.
\ee
A new operator that needs to be defined is that of the canonical momentum~\eqref{Pcanon2}
which is the Hamiltonian's main variable. For convenience we rewrite it in index notation,
\be
{\cal P}_{i}^k  =  p_{ic}^k + m_i  \epsilon_{k\ell m} x_{ic}^\ell  \cI_{m q}^{-1}  \epsilon_{qsn}  \sum_j x_{jc}^s p_{jc}^n.
\ee
The direct substitution recipe $ p_{ic}^k \to \hat{p}_{ic}^k$  produces a non-hermitian operator\footnote{For the conjugate
operator we find $(\hat{{\cal P}}_{i}^k)^\dagger =  \hat{p}_{ic}^k 
+ m_i  \epsilon_{k\ell m} \epsilon_{qsn}  \sum_j   \hat{p}_{jc}^n \left [ x_{ic}^\ell  \cI_{m q}^{-1} x_{jc}^s \right ]. $
The last term in this expression contains products that do not commute, e.g. $ \hat{p}_{ic}^n x_{ic}^n$ and  $ \hat{p}_{ic}^n \cI_{mq}^{-1}$. 
Therefore $(\hat{{\cal P}}_{i}^k)^\dagger \neq \hat{{\cal P}}_{i}^k$.}.
The proper method~\cite{Messiah} for defining the hermitian operator of a variable with a non-trivial dependence on $\mathbf{p}$ 
is to write the classical expression in a $\mathbf{p}$-symmetric form. In the case at hand this is,
\be
\bP_{i} = \mathbf{p}_{ic}  + \frac{1}{2} m_i \left ( \bOm \times \bx_{ic}  - \bx_{ic}  \times  \bOm \right ), 
\label{cPsymm}
\ee
and we define the corresponding operator as
\be
\hat{{\cal P}}_{i}^k  =  \hat{p}_{ic}^k + \frac{1}{2} m_i  \epsilon_{k\ell m} \epsilon_{qsn} \sum_j  x_{jc}^s
\left ( \, x_{ic}^\ell   \cI_{m q}^{-1}  \hat{p}_{jc}^n +  \hat{p}_{jc}^n [ x_{ic}^\ell  \cI_{m q}^{-1} ] \, \right ),
\label{cPhat2}
\ee 
where in the last term  $ x_{jc}^s$ has been factored out as it always commutes with $  \hat{p}_{jc}^n$ 
(unless $s=n$ in which case the term is zero). This operator is indeed hermitian:
\be
( \hat{{\cal P}}_{i}^k )^\dagger  
=  \hat{p}_{ic}^k + \frac{1}{2} m_i  \epsilon_{k\ell m} \epsilon_{qsn} \sum_j  x_{jc}^s
\left (\,    \hat{p}_{jc}^n [ x_{ic}^\ell   \cI_{m q}^{-1} ] +  x_{ic}^\ell  \cI_{m q}^{-1}  \hat{p}_{jc}^n \,\right ) = {\cal P}^k_i.
\ee

\subsection{The  `Mach-Schr\"odinger' equation}
\label{sec:MS}

The action of the canonical momentum operator on an arbitrary wavefunction $\Psi$ results in,
\be
\hat{{\cal P}}_{i}^k \Psi  = -i\hbar \left [  \nabla_{ic}^k \Psi + \frac{1}{2} m_i  \epsilon_{k\ell m} \epsilon_{qsn} \sum_j  x_{jc}^s
\left ( \, x_{ic}^\ell   \cI_{m q}^{-1}  \nabla_{jc}^n \Psi +  \nabla_{jc}^n [ x_{ic}^\ell  \cI_{m q}^{-1} \Psi ] \, \right ) \right ].
\label{cPhat3}
\ee 
The calculation of $\hat{{\cal P}}_{i}^2 \Psi $ involves several lines of algebra and the final result is rather unwieldy. 
However, it can be simplified if we notice that 
\be
 \nabla^n_{ic} ( \cI^{-1}_{mq} ) \sim \frac{m_i x_{ic}}{I^2}  \ll1,
 \ee
where $I$ is the typical size of the system's total moment of inertia. Working to first order in the small parameter $I^{-1}$ 
we find the much simpler expression,
\begin{align}
\hat{{\cal P}}_{i}^2 \Psi &= -\hbar^2 \left [ \nabla^2_{ic}  + 2 m_i   \epsilon_{k l m} x_{ic}^l \cI_{mq}^{-1}  \Big  (  \epsilon_{qkn}  \nabla^n_{ic}
+   \epsilon_{qsn}  \sum_j x_{jc}^s \nabla_{ic}^k \nabla_{jc}^n  \Big ) \right ] \Psi + {\cal O} \left  ( \frac{m_i^2}{I^2} \right ).
\label{cP2}
\end{align}
The theory's `Mach-Schr\"odinger' equation is 
\be
\hat{\cH} \Psi = \left [  \sum_i  \frac{\hat{{\cal P}}_{i}^2}{2 m_i} + V (\bx_{ij} )  \right ] \Psi =  i\hbar \frac{d  \Psi}{dt},
\ee
where $\Psi = \Psi (\bx_{ij}, t) = \Psi (\bx_{ic}-\bx_{jc}, t)$ is the system's \emph{global} wavefunction. 
After inserting~\eqref{cP2} this becomes,
 \be
i\hbar \frac{d  \Psi}{dt} = -\hbar^2  \sum_i  \left [ \frac{ \nabla^2_{ic}}{2m_i}  
+ \epsilon_{k l m}  x_{ic}^l \cI_{mq}^{-1}  \Big  ( \epsilon_{qkn}  \nabla^n_{ic}
+   \epsilon_{qsn}  \sum_j x_{jc}^s \nabla_{ic}^k \nabla_{jc}^n   \Big ) \right ] \Psi + V \Psi.
\label{MS1}
\ee
This fully relational equation, which describes a system of interacting quantum particles in their CM frame, 
is the main new result of this paper.  It is markedly more complicated than the standard  multi-particle Schr\"odinger equation 
in inertial space~\cite{Messiah} and, unlike that equation, it may not be separable.   
Our equation  bears some resemblance to the  non-inertial frame Schr\"odinger equation~\cite{Gulshani78} but fails to be equivalent 
to it due to the non-uniform function $\bOm(\bx_{jc},  \hat{\mathbf{p}}_{jc})$. As it was the case with the classical equation of motion, 
the new Machian terms are present even in the absence of an interaction potential, simply existing as a result of the presence of the 
bodies themselves. These terms however, considering that $I \gg m_i x_{ic}^2$ , are typically much smaller than the kinetic energy term.

The rest of the theory can be axiomatised as in standard QM. For instance, the expectation value 
$\langle \hat{Q} \rangle$ of any observable $Q$ is
\be
\langle \hat{Q} \rangle = \int dV \Psi^* \hat{Q} \Psi, \qquad dV = \prod_{i=1}^N d^3\bx_{ic}.
\ee
where the integral is to be evaluated in the full $N$-dimensional space of the system.

The hermitianity of $\hat{\cH}$ together with $i \hbar d\Psi/dt = \hat{\cH} \Psi$ leads to the textbook time evolution law
\be
i \hbar \frac{d \langle \hat{Q} \rangle}{dt} = \langle [\hat{Q}, \hat{\cH} ] \rangle.
\label{aver1}
\ee
With the help of this formula and the commutation relations (accurate to ${\cal O} \left (I^{-1} \right )$),
\begin{align}
[ \, x^k_{jc}, \hat{{\cal P}}^n_{i} \,] &= i \hbar \left [ \delta_{kn} \delta_{ij} 
+ m_i \epsilon_{nlm}  \epsilon_{qsk} \cI^{-1}_{mq} x_{jc}^s x_{ic}^l \right],  
\\
\nn \\
[ \, \hat{p}^k_{jc}, \hat{{\cal P}}^n_{i} \,] &= - \frac{i  \hbar}{2} m_i \cI_{mq}^{-1}  \Big [ \epsilon_{nlm} \epsilon_{qkb}
 \left (x^l_{ic} \hat{p}^b_{jc} + \hat{p}^b_{jc} x_{ic}^l \right )
+ 2 \delta_{ij} \epsilon_{nkm} \epsilon_{qsb} \sum_{j^\prime} x_{j^\p c} \hat{p}^b_{j^\p c} \Big ]
\nn \\
& =  - \hbar^2 m_i \cI_{mq}^{-1}  \Big [  \epsilon_{nlm} \epsilon_{qkb} x_{ic}^l \nabla_{jc}^b
+  \delta_{ij} \epsilon_{nkm} \epsilon_{qsb}\sum_{j^\p} x_{j^\p c}^s \nabla_{j^\p c}^b 
\nn \\
&\quad  + \frac{1}{2} \delta_{ij} \left ( \delta_{mq} \delta_{nk} - \delta_{mk} \delta_{nq} \right ) \Big ],
\end{align}
we can establish the validity of the Ehrenfest theorem in the new formulation. For example, setting
$\hat{Q} = x_j^k$ 
\be
i \hbar\frac{d \langle x_{jc}^k \rangle}{dt} = \langle [x_{jc}^k, \hat{\cH} ] \rangle 
 = \sum_i \frac{1}{2m_i} \langle [\,  \hat{x}_{jc}^k, \hat{{\cal P}}^2_{ic} \,] \rangle + \langle [  \, \hat{x}_{jc}^k, V \, ] \rangle.
\ee
For the commutators we have,
\be
 [  \, \hat{x}_{jc}^k, V (x_{ij})  \, ] =0, \qquad  
 [\,  \hat{x}_{jc}^k, \hat{{\cal P}}^2_{i} \,] =  [\,  \hat{x}_{jc}^k, \hat{{\cal P}}^n_{i} \,]  \hat{{\cal P}}^n_{i} 
 +  \hat{{\cal P}}^n_{i}  [\,  \hat{x}_{jc}^k, \hat{{\cal P}}^n_{i} \,],
\ee
and we find
\begin{align}
\frac{d \langle x_{jc}^k \rangle}{dt} &= \frac{\langle \hat{{\cal P}}^k_{jc} \rangle}{m_j} 
+ \frac{1}{2} \Big  \langle \epsilon_{ksq} x_{jc}^s \cI_{mq}^{-1}  \epsilon_{mln} \sum_i x_{ic}^l \hat{p}_{ic}^n
- \epsilon_{kqs} x_{jc}^s \epsilon_{mln} \cI_{mq}^{-1}  \sum_i x_{ic}^l \hat{p}_{ic}^n \Big \rangle. 
\end{align}
This result indeed coincides with the averaged value of the classical expression~\eqref{cPsymm}. 
A similar but significantly longer calculation establishes the same result for the classical equation of motion.

\section{Concluding remarks}
\label{sec:conclusions}

The new physics of the relational Schr\"odinger equation~\eqref{MS1} lies in its non-local character. 
This is a striking property not present in the non-inertial frame formulation of the standard Schr\"odinger 
equation~\cite{Gulshani78}. Even in the case of a vanishing interaction potential, each body is simultaneously 
affected by the rest of the system. Although the non-local terms are expected to be tiny compared to the usual 
kinetic terms they are, nevertheless, always present, thus forbidding (at least in a generic sense) the separability 
of the multi-body equation into decoupled single-body equations.  

It should be noted that the relational QM formulated in this paper is distinct from the more well known 
`relational QM'  of Ref.~\cite{Rovelli96}. In this latter theory standard QM is reinvented as a relational 
framework with respect to the observer-observed system pair and the concept of an observer's reality. 
In contrast, our framework is more `classical', replacing the standard absolute space  Schr\"odinger equation with 
a fully Machian equation. As pointed out earlier, our QM model is relational with respect to the bodies'
positions and momenta. 

Several different directions present themselves for future work. The non-locality of the model proposed in this
paper may have interesting ramifications for the non-local physics of entanglement and the EPR argument. 
Following the example of the classical mechanics model, the physics behind the Mach-Schr\"odinger equation 
could be unveiled with the help of a suitable `quantum bucket experiment' which should be designed to 
facilitate an exact solution of the aforementioned equation. We hope to be able to study some of these problems
in the near future.

\begin{acknowledgements}
The author is grateful to his transatlantic colleagues Daniel Kennefick, Aaron Johnson 
and Daniel Oliver at the Univ. of Arkansas for discussions that rekindled his interest in Machian physics 
and eventually led to this work. He also acknowledges support from a Fundaci\'on Seneca (Region de Murcia) 
grant No. 20949/PI/18. 
\end{acknowledgements}

\section*{Conflict of interest}
The author declares that he has no conflict of interest.



\end{document}